 \documentclass[pmlr,twocolumn,10pt]{jmlr} 

\usepackage{booktabs}
\usepackage{siunitx}

\usepackage[switch]{lineno}
\usepackage{hyperref}

\theorembodyfont{\upshape}
\theoremheaderfont{\scshape}
\theorempostheader{:}
\theoremsep{\newline}

\jmlrvolume{297}
\jmlryear{2025}
\jmlrworkshop{Machine Learning for Health (ML4H) 2025} 

 \title[Clinically-aligned Multi-modal Chest X-ray Classification]{Clinically-aligned Multi-modal Chest X-ray Classification}

\author{%
\Name{Phillip Sloan}\Email{phillip.sloan@bristol.ac.uk}\\
\Name{Edwin Simpson} \Email{edwin.simpson@bristol.ac.uk}\\
\Name{Majid Mirmehdi} \Email{majid.mirmehdi@bristol.ac.uk}\\
\addr University of Bristol, United Kingdom
}

\usepackage{amsmath}
\usepackage{graphicx}
\usepackage{amsfonts}
\usepackage{amssymb}
\usepackage[table]{xcolor}
\usepackage{color}

\newcommand{\framework}{CaMCheX\xspace}

\newcommand{\ps}{\textcolor[rgb]{0,0.0,0}} 

\ExplSyntaxOn
\NewDocumentCommand{\longdash}{ O{2} }
 {
  --\prg_replicate:nn { #1 - 1 } { \negthinspace -- }
 }
\ExplSyntaxOff

\begin{document}

\maketitle

\begin{abstract}
Radiology is essential to modern healthcare, yet rising demand and staffing shortages continue to pose major challenges. Recent advances in artificial intelligence have the potential to support radiologists and help address these challenges. Given its widespread use and clinical importance, chest X-ray classification is well suited to augment radiologists workflows. However, most existing approaches rely solely on single-view, image-level inputs, ignoring the structured clinical information and multi-image studies available at the time of reporting. In this work, we introduce \framework, a multimodal transformer-based framework that aligns multi-view chest X-ray studies with structured clinical data to better reflect how clinicians make diagnostic decisions. Our architecture employs view-specific ConvNeXt encoders for frontal and lateral chest radiographs, whose features are fused with clinical indications, history and vital signs using a transformer fusion module. This design enables the model to generate context-aware representations that mirror the reasoning in clinical practice. Our results exceed the state of the art 
for both the original MIMIC-CXR dataset and the more recent CXR-LT benchmarks, and highlight the value of clinically grounded multimodal alignment for advancing chest X-ray classification.

\end{abstract}

\paragraph*{Data and Code Availability}
The primary dataset used in this study was MIMIC-CXR \citep{MIMIC} with its CXR-LT 2023/2024 extensions \citep{Holste2024,CXR-LT2024}. For pretraining, we incorporated additional public datasets, including CheXpert \citep{CheXpert}, ChestXray14 \citep{Wang2017}, and VinDR \citep{VinDR}, while MIMIC-IV-ED \citep{MIMICIV} was leveraged to provide vital signs. Code is available here: \url{https://github.com/phillipSloan/CaMCheX}.

\paragraph*{Institutional Review Board (IRB)}
This work used public, deidentified imaging datasets with no direct human subject involvement; as such, IRB approval was not required.
\section{Introduction}
\label{sec:intro}

\begin{figure}[t]
\centering
\includegraphics[width=1.1\linewidth]{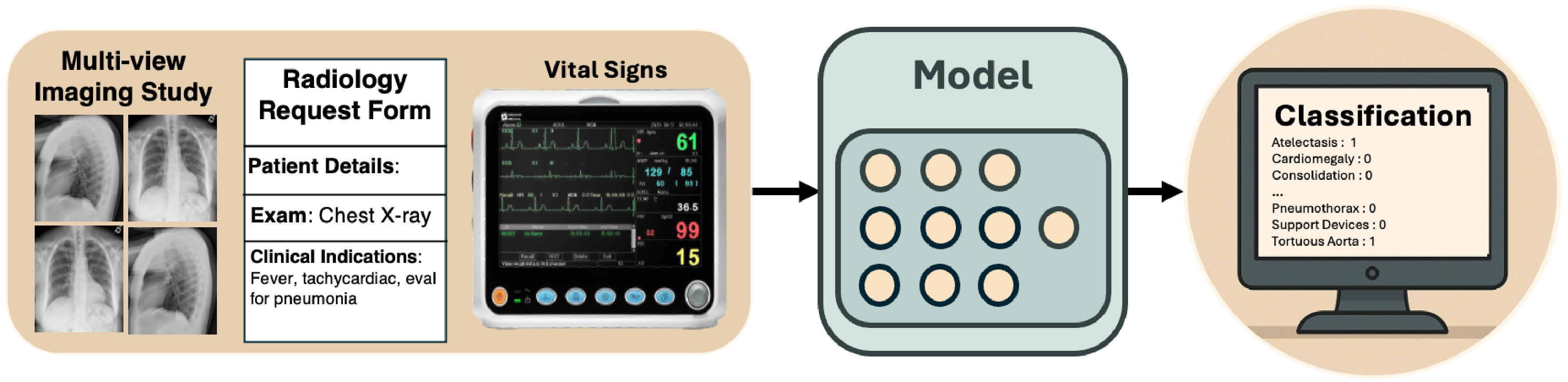}
\caption{{\textbf{Conceptual visualisation of \framework's Approach --}} \framework mirrors clinical decision-making by incorporating imaging data with complementary information such as clinical indications from radiology request forms and vital signs from A\&E triage, enabling accurate and clinically relevant chest X-ray classification.}
\label{fig:conceptual_framework}
\end{figure}

Radiology plays a vital role in modern healthcare, functioning as a key diagnostic tool for identifying diseases, tracking treatment response, and informing patient outcomes. Among radiological procedures, the chest X-ray is the most frequently performed examination and is crucial for the early detection of serious conditions such as pneumonia, lung cancer, and congestive heart failure. However, the increasing gap between radiology workforce capacity and service demand is a growing concern, with nearly all (99\%) clinical directors expressing worries about its impact on the safety and quality of care \citep{RCR}. Fatigue-related errors are an additional concern, potentially compromising diagnostic accuracy \citep{Stec2018}. 
In light of these challenges, there is an urgent need for innovation, and AI-based approaches to chest X-ray classification have been identified as a means to augment radiologists’ reporting \citep{RCR2}, with multimodal approaches highlighted as essential to maximise impact \citep{Banerji2025}.  


{Medical image classification presents a distinct set of challenges that set it apart from conventional image classification tasks. {For instance,} chest X-ray studies often comprise multiple views, including frontal (posteroanterior or anteroposterior) and lateral {(left or right)} projections, each offering distinct but complementary diagnostic insights.} 
Despite this inherent multi-view structure,
the majority of classification methods, such as those proposed in \cite{Nguyen-Mau2023, Jeong2023, Park2023, Wang2024GazeGNN, Dai2024UniChest, Wang2025CrossDomain}, approach the task at single image level, treating each chest X-ray independently. While this has lead to measurable progress, it overlooks the clinical reality 
{that radiologists interpret typically at the study level, where multiple views are reviewed together. Ignoring such anatomical and pathological context across views at study-level may result in missed findings and inaccurate diagnoses.}


\ps{Medical imaging tasks such as stroke outcome prediction \cite{Samak2022} and stroke lesion segmentation \cite{Samak2025} integrate spatial information across multiple planes or slices. These volumetric methods share conceptual parallels with the multi-view setting, as both combine multiple observations from the same patient. The distinction is that multi-view approaches operate on 2D chest X-rays rather than 3D volumes. Images may include frontal and lateral projections, which provide spatial context, or multiple images from the same perspective, each offering complementary diagnostic information.}

{Another limitation in current chest X-ray classification methods is the underuse of complementary clinical information by most existing approaches \citep{Kim2023, Nguyen-Mau2023, Jeong2023}. Clinical indications {are readily available at the time of reporting} and have been shown to enhance diagnostic accuracy in radiological practice \citep{Castillo2021}, and vital signs provide essential physiological context \citep{Candel2022, Rohmetra2023}, offering cues that can inform image interpretation. For example, an elevated temperature may prompt radiologists to focus more closely on features suggestive of pneumonia. Such structured data have been successfully incorporated into deep learning models in other medical domains to support outcome prediction \citep{Samak2023TranSOP, Hyland2024}. Although \cite{Jacenkow2022} incorporate clinical indications, their method remains limited to image-level classification, without study-level, multi-view reasoning or the inclusion of vital signs. To our knowledge, no existing chest X-ray classification approach jointly models both multi-view radiographs and structured clinical information, as found in datasets such as MIMIC-CXR \citep{MIMIC} and challenge benchmarks like CXR-LT 2023 \citep{Holste2024} {and CXR-LT 2024 \citep{CXR-LT2024}}. 



Other long-standing challenges, such as class imbalance and the multi-label nature of chest X-ray datasets, are well recognised in the field \citep{Kim2023, Holste2024}. These are commonly addressed using custom loss functions \citep{Kim2023, Park2023} or model ensembling \citep{Nguyen-Mau2023, Jeong2023, Hong2023}. We adopt a weighted asymmetric loss function to account for these factors, however, our primary focus is on enhancing multi-view integration and aligning model design more closely with clinical reasoning.



{In this paper, we address these limitations by introducing our \framework (Clinical-aligned Multi-modal Chest X-ray), \ps{a model that simultaneously integrates clinical indications and vital sign data with multi-view radiological images at the study level, advancing multimodal learning for chest X-ray classification.}  

\framework} incorporates a multi-view image backbone with separate, view-specific ConvNeXt encoders that independently process frontal and lateral images, enabling each encoder to specialise in the distinct characteristics of its respective view and extract complementary features. {These radiographic features are then integrated with structured clinical data through a transformer-based fusion module,}
{facilitating context-informed fusion that improves the accuracy and robustness of classification.}} {Figure~\ref{fig:conceptual_framework} presents a conceptual overview of our approach.}

{The design of our framework is grounded in feedback from five clinicians 
who confirmed they routinely review clinical indications and observations alongside available imaging.} 
Our contributions can be summarised as follows: (i) We propose \framework which integrates multi-view chest X-ray images with structured clinical data, including clinical indications and vital signs, in a manner that more closely reflects the clinical process followed by clinicians \citep{Banerji2025, Nensa2025}. (ii) The proposed method is capable of handling any number of frontal and lateral images using separate, view-specific encoders, and conditionally integrates clinical indications and vital signs when available, enabling robust performance across incomplete or heterogeneous inputs such as those in the MIMIC-CXR dataset. (iii) We perform extensive experiments and ablations to evaluate \framework against {state of the art models \citep{Kim2023, Li2024a},
CXR-LT 2023 \citep{Holste2024}{, CXR-LT 2024 \cite{CXR-LT2024}} and MIMIC-CXR \citep{MIMIC} label sets,
reporting improvements of 20.4\% in mean average precision (mAP) and 6.6\% in area under the receiver operating characteristic curve (AUROC) for the CXR-LT 2023 benchmark and an improvement of 12.8\% in AUROC for the original MIMIC-CXR labels.} 

\section{Related Works}\label{sec:relatedworks}


{Chest X-ray classification is the automated process of identifying and labeling findings in chest radiographs using machine learning models \citep{Chehade2024LungXrayReview}. It is typically formulated as a multi-label classification task, where a single image may exhibit multiple conditions such as pneumonia, cardiomegaly, or pleural effusion. This task supports clinical decision-making by assisting radiologists in detecting pathologies more efficiently and consistently.}
We focus our review on the under-utilisation of multi-view imaging \citep{Rubin2018, Zhu2021, Kim2023, CXR-LT2024} and on models that integrate multimodal information into chest X-ray classification \citep{Malik2024, Ketabi2023, Jacenkow2022}. {Several of the approaches that we discuss were developed for the CXR-LT 2023 \citep{Holste2024} and CXR-LT 2024 \cite{CXR-LT2024} challenges. While 2023 entries have separate publications, the 2024 submissions have not appeared independently and are therefore referenced by the challenge summary paper \citep{CXR-LT2024}}.

\subsection{Multi-view Chest X-ray Classification}
Although most chest X-ray classification methods generate predictions at the image level \citep{Nguyen-Mau2023, Jeong2023, Park2023, Hong2023, Verma2023, Yamagishi2023, KimC2023, Seo2023, Wang2024GazeGNN, Dai2024UniChest, Li2024a, Wang2025CrossDomain, CXR-LT2024}, this approach departs from clinical practice, where diagnoses are made at the study level by reviewing multiple views. However, several works have begun to explore multi-view approaches. \cite{Rubin2018}, \cite{Zhu2021}, and \cite{agostini2024} implemented architectures to handle frontal and lateral images separately, a limitation of these works is that the architecture adopts a fixed structure {and} assumes that a {frontal and lateral x-ray} are available and their designs do not accommodate multiple images from the same view. These assumptions are not reflective of the clinical landscape, or of MIMIC-CXR \citep{MIMIC}. To address this limitation, {competitors of CXR-LT 2023 \citep{Kim2023} and CXR-2024 \citep{CXR-LT2024} developed multi-view classification frameworks to aggregate features across views. While some of these approaches \cite{Kim2023} are invariant to the number {and view} of {the} input images, they employ shared encoders for frontal and lateral views which may limit the model's ability to capture the distinct anatomical and contextual features specific to each perspective. Our proposed \framework model builds upon these approaches by employing separate ConvNeXt encoders for frontal and lateral images, enabling the extraction of fine-grained, view-specific features.}


\subsection{Multi-modal Chest Diagnostics} 

In medical image classification, multimodality often refers to the combination of imaging types such as CT, MRI, and X-ray \citep{Li2024}. Other additional clinical data are also increasingly used \citep{Samak2023TranSOP, Samak2025StrokeReview, Hyland2024}, however such  additional modalities are sometimes not available at the time of initial presentation, 
while clinical indications and triage data are routinely available alongside the chest X-ray in an emergency setting, making them more immediately accessible for diagnostic support. 
However, to our knowledge, no existing approach combines study-level chest X-ray representations with both clinical indications and vital signs for classification. Several recent works have explored multimodal approaches for chest X-ray classification. \cite{Malik2024} proposed a multimodal framework combining chest X-rays, CT scans, and \ps{recordings of} cough sounds. While innovative, this diverges from clinical practice, where reporting clinicians typically lack direct patient contact and thus have no access to auditory cues. \cite{Ketabi2023} introduced a model incorporating chest X-rays, radiology reports, and eye-tracking data to improve classification and explainability. Although this demonstrates the value of textual information, it relies on radiology reports, which are produced post hoc and are unavailable at the point of initial review. \cite{Jacenkow2022} and \cite{Shurrab2024} included some clinical features but their models are limited to image-level predictions and do not incorporate contextual inputs such as vital signs, which offer further diagnostic value. \ps{\cite{Khader2023} developed a transformer-based architecture that integrates chest X-ray images with patient vital signs to improve classification performance. However, their approach remains limited to image-level predictions and does not incorporate clinical indications.}



{The proposed \framework integrates multi-view radiographs with structured clinical data, including clinical indications/history from request forms and vital signs which would be recorded at triage. This reflects real clinical workflows \cite{Banerji2025}, where radiologists consider both imaging and context at the study level. \framework enhances classification by aligning with how decisions are made in practice.}

\section{Methodology}\label{sec:methodology}
Our proposed \framework model integrates multi-view chest radiographs with structured clinical data. 
The training follows {a two-stage approach}. First, we fine-tune a ConvNeXt \citep{Liu2022} feature extractor for image level chest X-ray classification using noisy student training \citep{Xie2020}, establishing strong visual backbones. In the second stage, this fine-tuned ConvNeXt is incorporated into \framework to encode images as part of a study-level model that fuses radiographic features with clinical indications and vital signs observations. \\

\subsection{Stage 1: Fine-tuning ConvNeXt with Noisy Student Training } 

{In the first stage of our pipeline, we fine-tune a ConvNeXt model \citep{Liu2022}, using image-level chest X-ray classification and noisy student training. This prepares the model to act as a general purpose image feature extractor, producing representations across all projections before it is further adapted to view-specific encoding during the second stage training of \framework.}
{By training on all views, we enable the model to serve as a versatile image feature extractor, capable of producing representations irrespective of the projection before it is fine-tuned further to specific views during Stage 2. 

Following \cite{Kim2023}, we modify the ConvNeXt architecture to better accommodate the long-tailed, multi-class, multi-label nature of chest X-ray classification by replacing the standard global average pooling and linear classification head with an ML-Decoder \citep{Ridnik2023}, which allows the model to attend to diverse features when identifying different pathologies.} 

The process can be more formally stated as follows. Given a chest X-ray $\mathbf{x} \in \mathbb{R}^{C \times H \times W}$, where $C$ denotes the number of channels and $H \times W$ the spatial resolution, the ConvNeXt model produces a feature representation $\mathbf{z} \in \mathbb{R}^{C' \times H' \times W'}$. Here, $C'$, $H'$, and $W'$ correspond to the number of output feature channels and the reduced spatial dimensions resulting from convolutional processing. These feature maps are then passed to the MLDecoder classification head, which aggregates the spatial features and generates a prediction vector $\hat{\mathbf{y}} \in \mathbb{R}^{Q}$, where $Q = 26$ corresponds to the number of diagnostic labels defined by the CXR-LT label dataset. These predictions are trained and evaluated against the ground-truth labels $\mathbf{y}$ provided in the CXR-LT dataset \citep{Holste2024}.

To further enhance the quality of the extracted features, we adopt noisy student training \citep{Xie2020}, a method that leverages additional data to improve the initial training process. A teacher model is first trained on the available CXR-LT 2023 labels \citep{Holste2024} and is subsequently used to generate pseudo-labels for the \ps{CheXpert \citep{CheXpert}, ChestX-ray14 \citep{Wang2017} and VinDR \citep{VinDR} datasets, which are partly labelled, as discussed in Section~\ref{subset:datasets}.} A student model is subsequently trained on the merged set of labelled and pseudo-labelled data which then becomes the teacher for the next iteration, allowing for iterative refinement. In accordance with \cite{Xie2020}, we perform four such iterations, beyond which improvements become marginal. 

From the jointly trained model, we then train two dedicated encoders, $E_{F}$ and $E_{L}$, on frontal and lateral chest X-rays, respectively. Each is fine-tuned on its corresponding subset of the MIMIC-CXR dataset, enabling the model to adapt its general visual representations to the distinct anatomical and contextual features present in each view. This view-specific training supports more accurate and granular feature extraction, which is essential for effective multi-view classification within the \framework architecture.


\begin{figure*}[t]
\centering
\includegraphics[width=\textwidth]{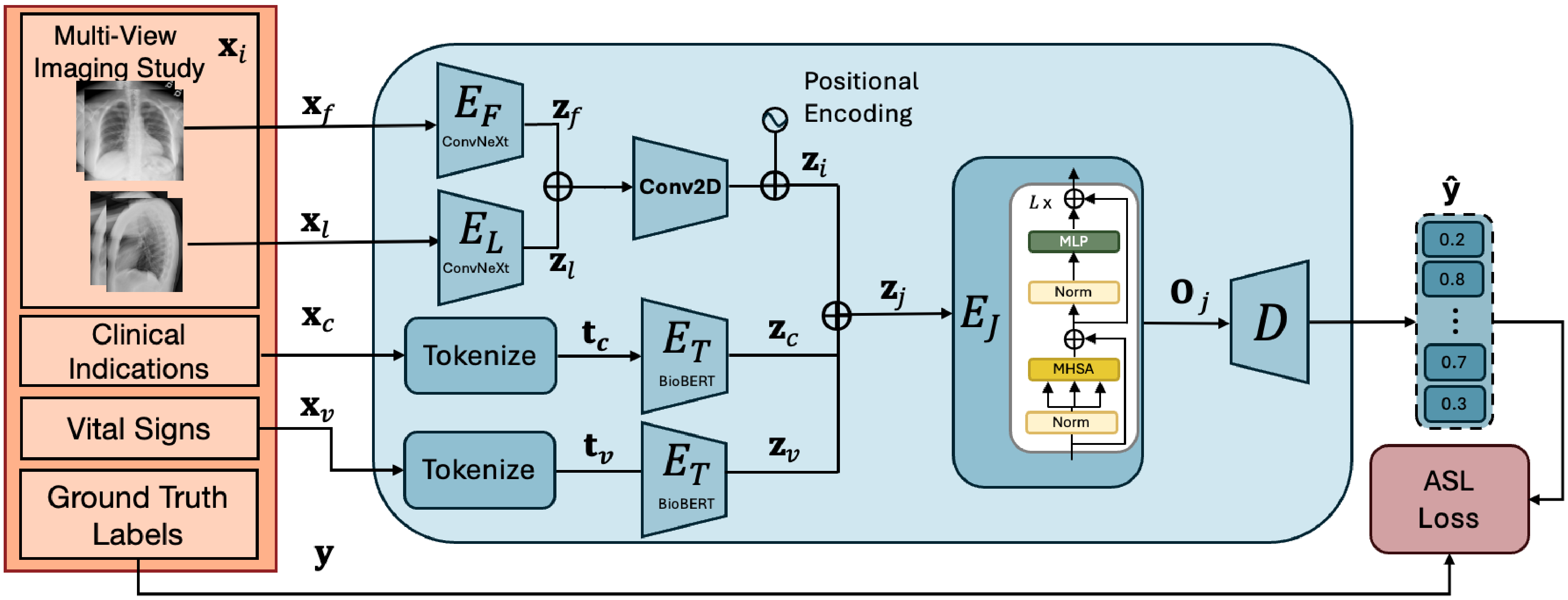}
\caption{Overview of the proposed \framework architecture (Stage 2). The model processes a set of multi-view images and its associated clinical indications and vital observations to classify chest abnormalities. Features extracted from each modality are distinguished using positional and segment embeddings, concatenated, and fused via a transformer-based fusion module $E_{j}$. The resulting joint representation is then passed to a classification head $D$ to produce multi-label predictions $\hat{\mathbf{y}}$.}
\label{fig:encoder_architecture}
\end{figure*}

\subsection{Stage 2: Multi-modal, Multi-view Feature Extraction}
Figure \ref{fig:encoder_architecture} illustrates the architecture of \framework. Following initial training at the image level, the second phase of our approach performs study-level classification by aggregating multiple chest X-rays per study.
We deploy the encoders ${E_{F}}$ and ${E_{L}}$ trained in Stage 1, while for our textual encoder, ${E_{T}}$, we use BioBERT \citep{Lee2020}, a BERT-based architecture pretrained on various medical texts. This domain-specific pretraining enables BioBERT to capture medical terminology and contextual nuances more effectively than standard BERT models, making it well-suited for extracting clinically relevant features from textual indications.

Consider a set of $N$ frontal ($f$) and lateral ($l$) X-ray images $\mathbf{x}_{i} \in \mathbb{R}^{N \times C \times H \times W}$, split into $\mathbf{x}_{f} \in \mathbb{R}^{f \times C \times H \times W}$ and $\mathbf{x}_{l} \in \mathbb{R}^{l \times C \times H \times W}$ respectively, based on view metadata, which is available in MIMIC-CXR \citep{MIMIC}. 
Each subset is processed by its respective ConvNeXt feature extractor, yielding $\mathbf{z}_f = {E_F}(\mathbf{x}_f)$ and $\mathbf{z}_l = {E_L}(\mathbf{x}_l)$, where $\mathbf{z}_{f} \in \mathbb{R}^{f \times C^{'} \times H^{'} \times W^{'}}$ and $\mathbf{z}_l \in \mathbb{R}^{l \times C' \times H' \times W'}$. 
The resulting features are merged into a joint tensor 
which is then passed through a shared convolutional layer to produce spatially reduced maps $\mathbf{z}_{i} \in \mathbb{R}^{N \times C^{'} \times H^{''} \times W^{''}}$. Positional encodings are then added to preserve spatial information within the transformer.

As $N$ varies between studies (within MIMIC-CXR, $1 \leq N \leq 11$, with $0 \leq f \leq 10$ and $0 \leq l \leq 5$), we cap the number of images per study at $N = 4$ for efficiency and memory constraints. For those with more than four, a random subset is sampled each epoch to preserve computational efficiency, which only impacts 191 out of 227,827 studies in MIMIC-CXR. 

The structured clinical inputs, clinical indications ($\mathbf{x}_{c}$) and vital signs ($\mathbf{x}_{v}$), are independently processed by the BioBERT encoder ${E_T}$. Each input can be represented as a token sequence, $\mathbf{x}_a = [w_1, w_2, \dots, w_S]$, which is tokend to $\mathbf{t}_a = [t_1, t_2, \dots, t_n]$ and passed through the encoder to obtain contextual representations: $\mathbf{z}_{a} = E_T(\mathbf{t}_{a}) \in \mathbb{R}^{n \times d}$. To integrate with the image features, the first token embedding $\mathbf{z}_a^{\text{cls}} \in \mathbb{R}^{d}$ is extracted and reshaped to match the spatial dimensions of the image feature maps, yielding $\mathbf{z}_a^{\text{spatial}} \in \mathbb{R}^{1 \times C^{'} \times H^{''} \times W^{''}}$. This process generates the features $\mathbf{z}_{c}$ and $\mathbf{z}_{v}$ for clinical indications and vital signs, respectively.

Once encoded segment embeddings are applied to allow the transformer to be able to distinguish between the generated features $\mathbf{z}_{i}$, $\mathbf{z}_{c}$ and $\mathbf{z}_{v}$, they are then concatenated along the sequence dimension to form $\mathbf{z}_{j} \in \mathbb{R}^{(N+2) \times C' \times H'' \times W''}$. 
The tensor is then 
passed through a transformer encoder ${E_{J}}$, producing the joint output, 
$\mathbf{O}_{j}$, which our classification head, ${D}$, decodes to generate a prediction vector $\hat{\mathbf{y}} \in \mathbb{R}^{Q}$. 
\ps{For studies missing clinical indications or vital signs, placeholders are used during both training and inference which allows the model to maintain a consistent input structure while preventing the introduction of synthetic or biased values. Example inputs to the model can be seen in Table~\ref{tab:study_inputs_examples}.}



\section{Experiments}\label{sec:experiments}
{Within this work, MIMIC-CXR \citep{MIMIC} with the CXR-LT 2023 \citep{Holste2024} labels serves as the primary dataset to fine-tune and evaluate our model, while several additional datasets are used during noisy student training to increase the number of training samples available. We also make use of the MIMIC-IV \citep{MIMICIV} dataset to extract patient vital signs from the available electronic hospital records. {While we perform supervd training with the CXR-LT labels, we report evaluatation results for \framework on both the MIMIC-CXR and CXR-LT label sets.}} 

\subsection{Datasets}\label{subset:datasets}
{We use several datasets in this work.} \textbf{MIMIC-CXR}\citep{MIMIC} is the largest available chest x-ray classification dataset, containing 377,110 radiographic images from 227,835 radiology studies conducted on 65,379 individuals presenting to the Beth Israel Deaconess Medical Center Emergency Department. Each study has an associated radiology report {containing the following sections: indication, technique, comparison, findings, and impression.} \ps{The clinical indications used in our model are extracted from the indication section of the corresponding radiology reports using the NLP extraction tool provided by \cite{MIMIC}, while diagnostic labels for training and evaluation are derived from the same reports using the CheXpert rule-based labeler \citep{CheXpert}, which assigns 14 categories corresponding to common thoracic diseases and findings.} 

\cite{Holste2024} introduce a set of MIMIC-CXR classification labels comprising of 26 categories {and later expanded to 40 labels in 2024 \cite{CXR-LT2024}. As a newly established benchmark, it plays a key role in defining the current state of the art in chest X-ray image classification. {The CXR-LT 2024 benchmark defines three tasks, but we focus on task 2, long tailed classification on a small, manually annotated test set of the same 26 labels as the CXR-LT 2023 benchmark.} \textbf{MIMIC-IV-ED} \citep{MIMICIV} is an emergency department dataset containing structured data from 441,561 visits by 322,872 patients again at Beth Israel Deaconess Medical Center. It includes triage observations such as temperature, heart rate, respiration rate and blood pressure, and can be linked to MIMIC-CXR via shared identifiers such as study-id and admission times. The overlap of patients between MIMIC-CXR and MIMIC-IV-ED is $61,856$.} 

We incorporate three additional chest X-ray datasets into our training pipeline. \textbf{CheXpert} \citep{CheXpert} consists of 224,316 chest X-ray images from 65,240 patients, labelled using a rule-based labeler developed for a medical image classification competition \citep{CheXpertCompetition} \ps{to extract 14 classification labels which can be considered positive, negative or uncertain. For Noisy Student training \citep{Xie2020} we retain positive and negative labels and treat uncertain or unmentioned cases as unlabelled}. {Building on CheXpert, CheXpert Plus \citep{CheXpertPlus} released the radiology reports for each study in \citep{CheXpert}, although, at present, these reports are only available for the training and development sets}. \textbf{NIH ChestX-ray14} \citep{Wang2017} contains 112,120 frontal chest X-ray images from 30,805 patients, each annotated with up to 14 disease labels. \textbf{VinDr-CXR} \citep{VinDR} includes 67,914 chest X-ray images collected from two hospitals in Vietnam, annotated with 15 abnormality labels\ps{, with Nodule and Mass findings merged into a single Nodule/Mass class for label consistency with the CXR-LT 2023 labels}. For all of these three datasets, we apply soft pseudo-labeling to any classes that do not overlap with the CXR-LT 2023 label set. {Neither NIH ChestX-ray14 \citep{Wang2017} nor VinDr-CXR \citep{VinDR} include associated clinical information.}

\subsection{Implementation details} 
Model implementation was developed in PyTorch and PyTorch Lightning, leveraging pretrained backbones from the timm and transformers libraries. Images were resized to $1024\times1024$.
Training was performed using the AdamW optimizer with a learning rate of $3\times10^{-5}$ and a weight decay of $1\times10^{-4}$ across all phases. To combat the long tail and label sparsity, we adopted a weighted asymmetric loss (ASL) \citep{BenBaruch2021} function. A cosine-annealing learning-rate schedule with a 5\% linear warm-up phase was used, with a batch size of 16. Exponential Moving Average (EMA) of model weights was used to stabilise learning. Training was performed on a single GH200 Superchip with 96Gb of vRAM. \ps{We repeat each experiment 3 times with different random seeds, reporting the mean in our results.}

In line with the CXR-LT benchmarks \citep{Holste2024, CXR-LT2024}, we evaluate \framework with commonly used metrics, mAP and AUC-ROC (macro). The primary evaluation metric is mAP since AUC-ROC can be disproportionately influenced due to long tailed class imbalance \citep{Holste2024} 
In contrast, mAP measures performance across decision thresholds and does not degrade under class imbalance \citep{Holste2024}. {For benchmarking against prior work, we also report results on the original 14-label MIMIC-CXR label dataset, where AUROC remains the standard metric of comparison.}

\begin{table*}[ht!]
\footnotesize
    \centering
    \resizebox{\textwidth}{!}{
    \begin{tabular}{lllcccc} \hline 
         \textbf{Model}   &  \textbf{Dataset}&\textbf{Year}&\textbf{Resolution}&\textbf{Architecture}& \textbf{mAP}& \textbf{AUROC} \\[0.5ex] \hline

 \rowcolor[gray]{0.95}
 \cite{Seo2023}&   CXR-LT 2023&2023&1024 & ResNet101 & 0.279 & 0.782 \\
 \cite{KimC2023}&   CXR-LT 2023 &2023&448 & TResNet50 & 0.328 & 0.822 \\
 \rowcolor[gray]{0.95}
  \cite{Yamagishi2023}&   CXR-LT 2023 &2023&224, 512 & EfficientNetV2 & 0.330 & 0.826 \\ 
 \cite{Verma2023} &  CXR-LT 2023 &2023 & 448 &ResNeXt101 \& DenseNet161 & 0.339 & 0.830\\
  \rowcolor[gray]{0.95}
   \cite{Hong2023}&   CXR-LT 2023 &2023&512 & ResNet50 & 0.349 & 0.836 \\
  \cite{Park2023}&   CXR-LT 2023 &2023&1024 & ConvNeXt & 0.351 & 0.838 \\  \rowcolor[gray]{0.95}
 \cite{Jeong2023}&   CXR-LT 2023 &2023&448 & ConvNeXt & 0.354 & 0.838\\ 
 \cite{Nguyen-Mau2023}&   CXR-LT 2023 &2023&512/768 & EfficientNetV2-S \& ConvNeXt & 0.354 & 0.836 \\ \rowcolor[gray]{0.95}
   CheXFusion \citep{Kim2023} &   CXR-LT 2023&2023&1024 & ConvNeXt \& Transformer & \underline{0.372} & \underline{0.850} \\
  {Ours (\framework)} &  CXR-LT 2023& 2025 & 1024 & ConvNeXt \& Transformer & \textbf{0.576} & \textbf{0.916}\\ \hline \rowcolor[gray]{0.95}
   Team F \citep{CXR-LT2024} & CXR-LT 2024 & 2024 & 384, 512 &ConvNeXt \& MaxViT-T & 0.509 & 0.829 \\
   Team E \citep{CXR-LT2024} & CXR-LT 2024 & 2024 & 336, 448, 512 &ViT-L & 0.511 & \underline{0.836} \\ \rowcolor[gray]{0.95}
   Team A \citep{CXR-LT2024} & CXR-LT 2024 & 2024 & 224, 384 &ConvNeXt & 0.519 & 0.834 \\
   Team C \citep{CXR-LT2024} & CXR-LT 2024 & 2024 & 1024 &ConvNeXt \& EfficientNet & \underline{0.526} & 0.833 \\ \rowcolor[gray]{0.95}
   {Ours (\framework)} &  CXR-LT 2024& 2025 & 1024 & ConvNeXt \& Transformer & \textbf{0.725} & \textbf{0.904}\\ \hline
   DualNet\ps{*} \citep{Rubin2018} & MIMIC-CXR & 2018 & - & DenseNet-121 & - & 0.721 \\ \rowcolor[gray]{0.95}
  {MMVM\ps{*}} \citep{agostini2024} & MIMIC-CXR & 2024 & 224 & ResNet VAE & - & 0.733 \\   
  \cite{Shurrab2024} &  MIMIC-CXR & 2024 & - & Siamese ViTs & - & 0.751 \\ \rowcolor[gray]{0.95}
   \cite{Yao2025} & MIMIC-CXR & 2025 & 224 & Instruction ViT & - & 0.782\\
  {MBRANet} \citep{Li2024} &  MIMIC-CXR & 2024 & 224 & ResNet50 & - & 0.801 \\  \rowcolor[gray]{0.95}
  MMBT \citep{Jacenkow2022} & MIMIC-CXR & 2022 & 224 & ResNet-50 \& BERT & - &  \underline{0.806}\\ 
  {Ours (\framework)} &  MIMIC-CXR & 2025 & 1024 & ConvNeXt \& Transformer & \textbf{0.669} & \textbf{0.934} \\ \hline
    \end{tabular}
    }
    \vspace{0.5cm}
    \caption{Quantitative evaluation with state of the art {approaches for the MIMIC-CXR}, CXR-LT 2023, and CXR-LT 2024 (Task 2) benchmarks. Best method's result is in bold, and second best is underlined. \ps{Results marked with * are reported on different MIMIC-CXR splits.}}
    \label{tbl:classification_results}
\end{table*}

\subsection{Results} 
Table~\ref{tbl:classification_results} compares the performance of our proposed \framework model against a range of recent approaches. 
{On the CXR-LT 2023 benchmark} 
our model achieves a mAP of 0.576 and an AUROC of 0.916, outperforming all competing methods by a substantial margin across both evaluation metrics. Compared to CheXFusion \citep{Kim2023}, which previously led the benchmark with an mAP of 0.372 and AUROC of 0.850, \framework demonstrates a relative improvement of 20.4\% and 6.6\% in mAP and AUROC respectively. Notably, several other top entries in the challenge, including those of \cite{Nguyen-Mau2023}, \cite{Jeong2023}, and \cite{Park2023} also rely on ConvNeXt backbones but do not match our performance, underscoring the benefit of our multimodal design rather than architectural choice alone. 

{On Task 2 of the CXR-LT 2024  benchmark, our model achieves a mAP of 0.725 and an AUROC of 0.904, establishing a new state of the art. The strongest competing entries for the CXR-LT 2024 benchmark are Team C and E \citep{CXR-LT2024} who scored highest on mAP and AUROC respectively. Team C employed a multimodal ensemble approach that combined multi-view, multi-scale image alignment with contrastive image-language pretraining, reaching an mAP of 0.526 and AUROC of 0.833. Team E used a foundational model, the DINOv2 model \citep{oquab2023dinov2}, along with multi-view, multi-resolution ensembling with a self-distillation loss, achieving 0.511 for mAP and 0.836 for AUROC. Compared to these approaches, \framework delivers an improvement of 19.9\% in mAP and 6.8\% in AUROC, underscoring the effectiveness of our multimodal design over the ensemble approaches within the CXR-LT 2024 competition.} 

When comparing against methods evaluated on the MIMIC-CXR labels, we see an improvement of 12.8\% over the state of the art \citep{Jacenkow2022}, demonstrating that our approach is generalisable beyond the CXR-LT setting, reinforcing the robustness of our multimodal design.\\
{As seen in Table~\ref{tbl:classification_mAP_longtail_results}, for the CXR-LT labels} \framework achieves the highest mAP on head classes, scoring 0.786, which significantly surpasses all baseline methods. The next best, CheXFusion \citep{Kim2023}, achieves 0.499, highlighting \framework’s strong ability to accurately detect the most common pathologies. For medium-frequency (body) classes, \framework {again substantially outperforms other methods},   achieving a mAP of 0.550. This result is particularly noteworthy, as it more than doubles the score of CheXFusion, which records 0.246. While some existing methods show strong head-class performance, they typically struggle in the medium range, whereas \framework demonstrates clear robustness. Finally, in the tail group, \framework achieves a mAP of 0.408, representing a marked improvement over previous methods such as \citet{Jeong2023}, who achieve an mAP of 0.246. This gain on rare pathologies underscores \framework’s ability to generalise effectively across the entire distribution, maintaining strong performance even on the most infrequent conditions.


Table \ref{tbl:pathology_results} presents per-pathology results on the CXR-LT 2023 benchmark, grouped by class frequency. Overall performance follows the expected long-tail trend, with head classes achieving the highest mean AP (0.786) due to their greater representation and more distinctive visual features. Body classes exhibit moderate performance (mean AP = 0.550), indicating that the model generalizes well even when fewer samples are available. Tail classes, while the most challenging (mean AP = 0.408), still maintain strong discriminative capability despite limited training data. Notably, Pleural Other remains a difficult category, reflecting both its rarity and the inherent ambiguity of its label. These results demonstrate that, although performance naturally correlates with class frequency, our model performs robustly across all groups. This consistent performance across the head, body, and tail distributions highlights that leveraging complementary information effectively enhances classification performance for both frequent and rare pathologies.

Altogether, these results demonstrate that \framework delivers consistently strong performance across all frequency groups, outperforming the state of the art in both common and rare pathology detection. 

\begin{table*}[t!]
\footnotesize
\centering
\begin{tabular}{llccc} \hline
    \textbf{Model}  &\textbf{Year}& \multicolumn{3}{c}{\textbf{mAP}} \\ \cline{3-5}
     && \textbf{Head (\textgreater10\%)} & \textbf{Body (1–10\%)} & \textbf{Tail (\textless 1\%)} \\[0.2ex] \hline
    \rowcolor[gray]{0.95}
    \cite{Seo2023}  &2023& 0.420 & 0.154 & 0.086 \\
    \cite{KimC2023}  &2023& 0.461 & 0.195 & 0.195 \\
    \rowcolor[gray]{0.95}
    \cite{Yamagishi2023}  &2023& 0.460 & 0.195 & 0.216 \\ 
     \cite{Verma2023}  &2023& 0.476 & 0.210 & 0.179 \\   

    \rowcolor[gray]{0.95}
    \cite{Hong2023}  &2023& 0.474 & 0.218 & 0.243 \\
    \cite{Park2023}  &2023& 0.480 & 0.221 & 0.220 \\
    \rowcolor[gray]{0.95}
    \cite{Jeong2023}  &2023& 0.477 & 0.226 & \underline{0.246} \\

    \cite{Nguyen-Mau2023}  &2023& 0.482 & 0.226 & 0.227 \\    
    \rowcolor[gray]{0.95}
    CheXFusion \citep{Kim2023}  &2023& \underline{0.499} & \underline{0.246} & 0.242 \\
    {Ours (\framework})  & - & \textbf{0.786} & \textbf{0.550} & \textbf{0.408} \\ \hline
\end{tabular}
\vspace{0.5cm}
\caption{Per-frequency-group mAP results (Head, Medium, Tail) for the CXR-LT {2023} benchmark. Best result in bold, second is underlined.}
\label{tbl:classification_mAP_longtail_results}
\end{table*}

\begin{table*}[t!]
\footnotesize
    \centering
    \begin{tabular}{lcccccc} \hline 
         \textbf{Model}   & \textbf{Images}& \textbf{Multi-View } &  \textbf{Clinical Indications } &\textbf{Vital Signs } & \textbf{mAP }& \textbf{AUROC} \\
          & &\textbf{(MV)}& \textbf{(CI)}&\textbf{ (VS)} &  &  \\ \hline
     \rowcolor[gray]{0.95}
    Single-View&  $\checkmark$ &&  && 0.346& 0.837\\
    \ps{CI only} &  & &  $\checkmark$ && \ps{0.268} & \ps{0.745} \\
    \rowcolor[gray]{0.95}
   \ps{VS only} & &&  &$\checkmark$ & \ps{0.128} & \ps{0.580} \\
    Multi-View&  $\checkmark$&$\checkmark$&  && 0.392& 0.859\\
        \rowcolor[gray]{0.95}
    MV + CI&  $\checkmark$&$\checkmark$&   $\checkmark$&& 0.551 & 0.903\\ 
 \framework& $\checkmark$& $\checkmark$& $\checkmark$&$\checkmark$& \textbf{0.576} & \textbf{0.916}\\ \hline
    \end{tabular}
    \vspace{0.5cm}
    \caption{{Ablation study illustrating the contribution of each modality to overall performance on the CXR-LT {2023} benchmark.}}
    \label{tbl:ablation_results}
\end{table*}

\subsection{Ablation} Table~\ref{tbl:ablation_results} presents an ablation study evaluating the contribution of each component on the CXR-LT 2023. Starting from a single-view baseline, we observe an improvement when moving to a multi-view setup, with a 3.6\% increase in mAP and 1.7\% increase in AUROC. This demonstrates the benefit of study-level classification over image-level, which provides complementary anatomical information. \ps{Examining the contribution of the non-imaging modalities in isolation, clinical indications achieve moderate performance of 0.268 mAP and 0.745 AUROC, suggesting clinical indications provide useful context but they are insufficient on their own to capture the full complexity of thoracic observations. Vital signs performs marginally above chance when used in isolation, with a mAP of 0.128 and an AUROC of 0.580, suggesting that measures such as heart rate are, on their own, weak predictors of thoracic findings. When integrated with imaging features, however, both modalities enrich the model's generated features.} 

It can be seen that incorporating clinical information into the multi-view model improves performance, improving mAP by 16.5\% in mAP and 4.4\% in AUROC, indicating that the inclusion of contextual clinical information supports more accurate predictions. Integrating vital signs achieves the highest performance, with \framework achieving a further improvement of 2.5\% in mAP and 1.3\% in AUROC. These results demonstrate that each modality, multi-view images, clinical indications and vital signs contribute to the models capabilities.

\section{Conclusions and Future Work}\label{sec:conclusion}

We introduce \framework, a multimodal framework for chest X-ray classification that integrates multi-view radiographs with structured clinical data, including clinical indications and vital signs. The design of our framework is grounded in feedback from five clinicians, who confirmed that they routinely review clinical context alongside both frontal and lateral chest radiographs as part of their diagnostic workflow. By aligning image features with structured clinical information through a transformer-based module, \framework more closely mirrors real-world diagnostic reasoning, where radiologists integrate diverse information sources to build a clinical picture that informs their decisions \citep{Nensa2025}. Our model employs view-specific encoders and a context-aware fusion mechanism, enabling it to capture both anatomical and physiological cues. 

Evaluated on {the MIMIC-CXR and} CXR-LT benchmarks, \framework achieves state of the art performance for both tasks. In addition to strong overall results {and transferability across datasets}, it demonstrates robust generalisation across head, body, and tail classes. These findings highlight the value of clinically grounded, multimodal models in advancing automated chest X-ray interpretation. {At present, clinical information such as indications and other demographics is not available for the test set of CheXpert Plus \citep{CheXpert} preventing direct validation of \framework against methods that adopt CheXpert as a baseline. If this becomes available it will allow a more rigorous evaluation against further models.}

\ps{Image resolutions have been shown to influence performance in chest X-ray classification \citep{Kim2023} and exploring the impact of resolution within our multimodal framework represents a valuable direction for future work.}
\ps{While \framework incorporates clinical information typically available to radiologists at the time of interpretation, such as vital signs and clinical indications, other context that informs diagnostic reasoning, such as patient demographics, were not available for this work due to dataset scope. Prior imaging is available within MIMIC-CXR and offers a valuable opportunity for longitudinal analysis, which we identify as an important direction for future work.}

Looking ahead, the \framework encoder can be integrated into automated radiology report generation (ARRG) systems, a field growing rapidly \citep{Sloan2025}. As this area expands, grounding generative models in clinically validated, multimodal features is increasingly important. Our classification results offer a robust foundation for future systems to generate accurate, context-aware, and clinically meaningful reports, aligning these ARRG systems with multimodal clinical information.

\section*{Acknowledgement}
The 1st Author wishes to acknowledge and thank the financial support of the UKRI (Grant ref EP/S022937/1) and the University of Bristol. The authors acknowledge the use of resources provided by the Isambard-AI National AI Research Resource (AIRR). Isambard-AI is operated by the University of Bristol and is funded by the UK Government’s Department for Science, Innovation and Technology (DSIT) via UK Research and Innovation; and the Science and Technology Facilities Council [ST/AIRR/I-A-I/1023].

\bibliography{jmlr}
\newpage
\appendix
\onecolumn
\renewcommand{\thetable}{A\arabic{table}}
\setcounter{table}{0} 
\section{Example Inputs}\label{apd:a}

\begin{table*}[ht!]
\footnotesize
\centering
\begin{tabular}{p{0.11\linewidth} p{0.07\linewidth} p{0.36\linewidth} p{0.42\linewidth}} \hline
\textbf{Study ID} & \textbf{Images} & \textbf{Clinical Indications} & \textbf{Vital Signs (text input)} \\[0.5ex] \hline
\rowcolor[gray]{0.95} 
50000766 & 2 & year old woman s/p right vats mediastinal bx and pericardial effusion pna pna . & Temperature: 97.9 \textbar{} Heart rate: 109.0 \textbar{} Respiratory rate: 22.0 \textbar{} O2 Saturation: 94.0 \textbar{} Systolic BP: 136.0 \textbar{} Diastolic BP: 54.0 \textbar{} Gender: F \\
50000801 & 3 & year old man with bilateral pleural effusions s/p drainage r/o ptx r/o ptx . & Temperature: 98.8 \textbar{} Heart rate: 70.0 \textbar{} Respiratory rate: 18.0 \textbar{} O2 Saturation: 100.0 \textbar{} Systolic BP: 99.0 \textbar{} Diastolic BP: 56.0 \textbar{} Gender: M \\
\rowcolor[gray]{0.95} 
50020872 & 2 & -year-old female with shortness of breath . & Temperature: 101.2 \textbar{} Heart rate: 115.0 \textbar{} Respiratory rate: 18.0 \textbar{} O2 Saturation: 100.0 \textbar{} Systolic BP: 116.0 \textbar{} Diastolic BP: 77.0 \textbar{} Gender: F \\
50173309 & 2 & year old woman with exertional dyspnea mild hypoxia eval for acute process . & Temperature: 98.5 \textbar{} Heart rate: 89.0 \textbar{} Respiratory rate: 18.0 \textbar{} O2 Saturation: 98.0 \textbar{} Systolic BP: 127.0 \textbar{} Diastolic BP: 74.0 \textbar{} Gender: F \\
\rowcolor[gray]{0.95} 
50286142 & 2 & hodgkins lymphoma atypical chest pain evaluation . & Temperature: 99.8 \textbar{} Heart rate: 130.0 \textbar{} Respiratory rate: 20.0 \textbar{} O2 Saturation: 96.0 \textbar{} Systolic BP: 84.0 \textbar{} Diastolic BP: 48.0 \textbar{} Gender: M \\
50349057 & 1 & -year-old female with history of chest pain left rib pain . & Temperature: 98.3 \textbar{} Heart rate: 68.0 \textbar{} Respiratory rate: 16.0 \textbar{} O2 Saturation: 99.0 \textbar{} Systolic BP: 135.0 \textbar{} Diastolic BP: 72.0 \textbar{} Gender: F \\ \hline
\end{tabular}
\vspace{0.5cm}
\caption{Example study-level inputs showing clinical indications and the vital-signs text are represented and processed by our model.}
\label{tab:study_inputs_examples}
\end{table*}
\clearpage  
\section{Pathology-Level results}\label{apd:b}

\begin{table*}[ht!]
\footnotesize
\centering
\begin{tabular}{lcc} \hline
\textbf{Pathology} & \textbf{AP} & \textbf{AUROC} \\[0.5ex] \hline
\multicolumn{3}{l}{\textbf{Head (\textgreater10\%)}} \\ \hline
\rowcolor[gray]{0.95} Atelectasis & 0.6569 & 0.8489 \\
Cardiomegaly & 0.6999 & 0.8426 \\
\rowcolor[gray]{0.95} Edema & 0.8275 & 0.9507 \\
Lung Opacity & 0.6512 & 0.8152 \\
\rowcolor[gray]{0.95} No Finding & 0.7409 & 0.9398 \\
Pleural Effusion & 0.8848 & 0.9509 \\
\rowcolor[gray]{0.95} Pneumonia & 0.8813 & 0.9451 \\
Support Devices & 0.9447 & 0.9674 \\ 
\hline
\rowcolor[gray]{0.9} \textbf{Mean (Head)} & \textbf{0.786} & \textbf{0.920} \\ \hline

\multicolumn{3}{l}{\textbf{Body (1-10\%)}} \\ \hline
\rowcolor[gray]{0.95} Calcification of the Aorta & 0.2174 & 0.9300 \\
Consolidation & 0.4987 & 0.8877 \\
\rowcolor[gray]{0.95} Emphysema & 0.2674 & 0.9274 \\
Enlarged Cardiomediastinum & 0.3011 & 0.7006 \\
\rowcolor[gray]{0.95} Fracture & 0.6807 & 0.9411 \\
Hernia & 0.7284 & 0.9585 \\
\rowcolor[gray]{0.95} Infiltration & 0.8976 & 0.9768 \\
Mass & 0.6413 & 0.9365 \\
\rowcolor[gray]{0.95} Nodule & 0.3883 & 0.8757 \\
Pneumothorax & 0.8827 & 0.9714 \\ 
\hline
\rowcolor[gray]{0.9} \textbf{Mean (Body)} & \textbf{0.550} & \textbf{0.911} \\ \hline

\multicolumn{3}{l}{\textbf{Tail (\textless 1\%)}} \\ \hline
\rowcolor[gray]{0.95} Fibrosis & 0.4428 & 0.9575 \\
Lung Lesion & 0.7221 & 0.9715 \\
\rowcolor[gray]{0.95} Pleural Other & 0.0537 & 0.9142 \\
Pleural Thickening & 0.1508 & 0.8906 \\
\rowcolor[gray]{0.95} Pneumomediastinum & 0.6579 & 0.9755 \\
Pneumoperitoneum & 0.4677 & 0.9645 \\
\rowcolor[gray]{0.95} Subcutaneous Emphysema & 0.6530 & 0.9924 \\
Tortuous Aorta & 0.1651 & 0.8760 \\ 
\hline
\rowcolor[gray]{0.9} \textbf{Mean (Tail)} & \textbf{0.408} & \textbf{0.943} \\ \hline
\end{tabular}
\vspace{0.5cm}
\caption{Per-pathology classification results on the \textbf{CXR-LT 2023} dataset grouped by Head, Body, and Tail categories depending on their an . MeAverage Precision (mAP) and Area Under the ROC Curve (AUROC) are reported for each pathology, along with group means.}
\label{tbl:pathology_results}
\end{table*}

\end{document}